\documentclass[]{pasj01}
\usepackage{mathpazo}
\usepackage[T1]{fontenc} 
\usepackage{lineno}

\Received{$\langle$reception date$\rangle$}
\Accepted{2023 September 29}
\Published{$\langle$publication date$\rangle$}

\begin{document}

\title{Gamma-Ray Emission in the Seyfert Galaxy NGC~4151: Investigating the Role of Jet and Coronal Activities}
\author{Yoshiyuki \textsc{Inoue}\altaffilmark{1,2,3}}
\altaffiltext{1}{Department of Earth and Space Science, Graduate School of Science, Osaka University, Toyonaka, Osaka 560-0043, Japan}
\altaffiltext{2}{Interdisciplinary Theoretical \& Mathematical Science Program (iTHEMS), RIKEN, 2-1 Hirosawa, Saitama 351-0198, Japan}
\altaffiltext{3}{Kavli Institute for the Physics and Mathematics of the Universe (WPI), UTIAS, The University of Tokyo, Kashiwa, Chiba 277-8583, Japan}
\email{yinoue@astro-osaka.jp}

\author{Dmitry 
\textsc{Khangulyan}\altaffilmark{4}}
\altaffiltext{4}{Graduate School of Artificial Intelligence and Science, Rikkyo University, Nishi-Ikebukuro 3-34-1, Toshima-ku, Tokyo 171-8501, Japan}
\email{d.khangulyan@rikkyo.ac.jp}


\maketitle

\begin{abstract}
NGC 4151, a nearby Seyfert galaxy, has recently been reported to emit gamma rays in the GeV range, posing an intriguing astrophysical mystery. The star formation rate of NGC~4151 is too low to explain the observed GeV flux, but the galaxy is known for its coronal activity in X-ray and jet activity in radio. We propose that either the combination of these two activities or the jet activity alone can account for the gamma-ray spectrum. An energy-dependent variability search will allow one to distinguish between the two scenarios, as the coronal component can only contribute at energies of $\lesssim1$ GeV. Our analysis also indicates that it might still be difficult to see coronal neutrinos from the apparently X-ray brightest Seyfert NGC~4151 with current-generation neutrino observatories. 
\end{abstract}

\section{Introduction}
The Fermi-Large Area Telescope (LAT) allowed the detection of 6658 sources \citep{Fermi2022ApJS..260...53A} with its 12-year survey data. Among these numerous GeV gamma-ray sources, a new gamma-ray population, Seyfert galaxies, has started to emerge. The importance of these new gamma-ray sources is enhanced by the reported detection of high-energy neutrinos from a Type-2 Seyfert galaxy NGC~1068. After the initial tentative detection, the follow-up studies revealed the growing significance of this detection, with its confidence level now being 4.2$\sigma$ \citep{IceCube2022Sci...378..538I}. The gamma-ray and neutrino production mechanism in Seyfert galaxies is an intriguing mystery. 

In the X-ray sky, the brightest Seyfert galaxy is NGC~4151 \citep{Oh2018ApJS..235....4O}. Because of its X-ray brightness, this object has been a primary target in searching for high-energy gamma-ray signals from Seyferts \citep{Bignami1979ApJ...232..649B, Lin1993ApJ...416L..53L, Fermi2012ApJ...747..104A}. Utilizing Fermi-LAT data with 11.1~yr long exposure, \citet{Ajello2021ApJ...921..144A} have already reported $4.2\sigma$ significance, with a best-fit photon index of $-1.9^{+0.5}_{-0.3}$ and a best-fit flux of $6.3^{+3.7}_{-3.8}\times10^{-11}~ \mathrm{ph\ cm^{-2}\ s^{-1}}$ in the 1--800 GeV band. Then, after 14~yr of the Fermi-LAT survey, \citet{Peretti2023arXiv230303298P} finally reported the detection of $5.5\sigma$ in the energy range of 0.1--1000~GeV. The reported best-fit photon index is $-2.39\pm0.18$, slightly softer than that reported by \citet{Ajello2021ApJ...921..144A}, with a best-fit normalization of $(1.3\pm0.2) \times 10^{-13}~ \mathrm{MeV^{-1}\ cm^{-2}\ s^{-1}}$ at the pivot energy $E_0=1~\mathrm{GeV}$. The resulting gamma-ray luminosity in the energy band of 0.1--100 GeV is $3.7\times10^{40}~\mathrm{erg~s^{-1}}$ for a source at the distance of $D=15.8~\mathrm{Mpc}$.

One immediate candidate for the origin of gamma-ray emission is star formation activity, as is commonly seen in starburst galaxies (see, e.g., \cite{Fermi2012ApJ...755..164A, Ajello2020ApJ...894...88A, Owen2021MNRAS.506...52O}). The star formation rate (SFR) of NGC~4151 is 0.251--0.95~$\mathrm{ M_\odot~yr^{-1}}$ based on the measurements of the H$ \alpha $ emission line \citep{Erroz-Ferrer2015MNRAS.451.1004E, Theios2016ApJ...822...45T}. Following the empirical correlation between SFR and gamma-ray luminosity \citep{Fermi2012ApJ...755..164A, Ajello2020ApJ...894...88A}, the expected SFR-originated GeV gamma-ray luminosity is (0.28--1.5)$\times10^{39}~\mathrm{erg~s^{-1}}$, which is insufficient to explain the observed gamma-ray emission.

\citet{Peretti2023arXiv230303298P} have proposed the wind termination
shock formed by continuous ultra-fast outflows (UFOs) as the possible gamma-ray production site (see e.g., \cite{Lamastra2016A&A...596A..68L, Peretti2023arXiv230113689P}). UFOs are the accretion disk winds with a velocity of $\gtrsim0.1c$ \citep{Tombesi2010A&A...521A..57T}. However, the existence of continuous UFO activity in NGC~4151 is uncertain. UFO activity in NGC~4151 was observed only in the 2006 XMM--Newton Obs ID 0402660201 \citep{Tombesi2010A&A...521A..57T, Couto2016ApJ...833..191C}. In addition, NGC~4151 has repeatedly shown changing-look (CL) phenomena (see, e.g., \cite{Yaqoob1989MNRAS.236..153Y, Ricci2022arXiv221105132R, Li2022ApJ...936...75L}), which show large variability amplitudes in both continuum and emission lines.  

NGC~4151 is also known as one of the brightest radio-quiet active galactic nuclei (AGNs) in the radio band, and it has an elongated radio jet structure \citep{Wilson1982ApJ...263..576W, Ulvestad1998ApJ...496..196U, Ulvestad2005AJ....130..936U, Williams2017MNRAS.472.3842W}. The extent of the jet is $\sim3\farcs5$, corresponding to $\sim270$~pc. This collimated outflow requires a continuous energy supply from the central black hole (BH) activity. This jet is another plausible gamma-ray production site in analogy with blazars and radio galaxies. Also, even a weak jet propagating through the interstellar medium can cause cosmic ray acceleration with power exceeding the limit set by the star formation activity \citep{2022ApJ...936L...1M}.

Finally, as NGC~4151 is the brightest Seyfert in the X-ray sky, the BH corona, where the intense X-ray emission is generated by thermal Comptonization, is another plausible gamma-ray production site \citep{Inoue2019ApJ...880...40I, Murase2020PhRvL.125a1101M, Gutierrez2021A&A...649A..87G, Eichmann2022ApJ...939...43E}. It should be noted that coronal gamma rays cannot contribute at $\gtrsim$ GeV due to attenuation by the pair production process.

In this Letter, we investigate the origin of the observed gamma-ray emission from the nearby Seyfert galaxy NGC~4151. We consider two scenarios: one in which the emission arises from a combination of jet and coronal activity and another in which the emission is solely due to the jet. We further discuss how to discriminate between these two scenarios. We also examine the expected neutrino signals from coronal activity.

\section{Observed Properties of NGC~4151}
NGC~4151 has a central BH mass of $2.1\times10^7~M_\odot$ based on reverberation mapping measurements \citep{DeRosa2018ApJ...866..133D}. The reported distance to NGC~4151 ranges from 4 to 29~Mpc (see e.g., \cite{deVaucouleurs1981ApJ...248..408D, Theureau2007A&A...465...71T, Honing2014Natur.515..528H,Yoshii2014ApJ...784L..11Y, Yuan2020ApJ...902...26Y}). In this paper, we take the latest reported distance of $D=15.8$~Mpc based on the Cepheid distance measurement \citep{Yuan2020ApJ...902...26Y}. This corresponds to an angular scale of $\sim76.7~\mathrm{pc\ arcsec^{-1}}$.

NGC~4151 has an elongated radio jet structure with more than six knots seen in radio \citep{Wilson1982ApJ...263..576W, Ulvestad2005AJ....130..936U, Williams2017MNRAS.472.3842W}. 
The jet has a viewing angle of $\approx40^\circ$ \citep{Pedlar1993MNRAS.263..471P, Robinson1994A&A...291..351R, Vila-Vilaro1995A&A...302...58V} and the jet speed is $\le0.04c$ \citep{Williams2017MNRAS.472.3842W}. Thus, the relativistic effects are weak in the jet of NGC~4151; in particular, we do not expect a noticeable beaming effect. 

Among the observed radio knots, the brightest knot is the core component, known as the C4 component, having an integrated flux density of $72.0\pm0.002$~mJy at 1.5~GHz from the eMERLIN measurement carried out in 2015 \citep{Williams2017MNRAS.472.3842W}, which shows about 50\% brightening compared to the 1993 MERLIN observation \citep{Mundell1995MNRAS.272..355M}. The size of the component is $161.2\times150.6~\mathrm{mas}^2$. Assuming spherical symmetry, the physical radius of C4 is $R\approx12$~pc. \citet{Williams2017MNRAS.472.3842W} derived a spectral index of $1.4$ for the radio components using the full 512~MHz bandwidth of eMERLIN, which is consistent with the wider band spectrum obtained earlier \citep{Carral1990ApJ...362..434C, Pedlar1993MNRAS.263..471P}. The magnetic field strength of the C4 component is estimated to be $B=9.6\times10^{-4}~\mathrm{G}$ based on the minimum energy condition (see Table. 1 in \authorcite{Williams2017MNRAS.472.3842W} \yearcite{Williams2017MNRAS.472.3842W}). Since Fermi-LAT started its survey in 2008, in our analysis we consider only eMERLIN data.

Regarding the disk component, the spectral type of NGC~4151 changes over time as it is a CL AGN. For example, it changed from type~1 to type~2 in the 1980s \citep{Penston1984MNRAS.211P..33P}, while \citet{Shapovalova2008A&A...486...99S} found a change from type~1.5 to type~1.8 between 1996 and 2006. \citet{Mahmoud2020MNRAS.491.5126M} have recently modeled the broadband thermal emission of the disk in NGC~4151 and its spectral evolution based on their spectral timing model \citep{Gardner2017MNRAS.470.3591G, Mahmoud2020MNRAS.491.5126M}. The model reproduces the Swift monitoring data covering optical / UV / X-ray with a $\sim6$~hour cadence for a total of 69~days in 2016 \citep{Edelson2017ApJ...840...41E}. Reprocessed emission by the dusty torus appears in the IR band, with a reported $K$-band magnitude of the nucleus component of 133~mJy (see e.g., \cite{Kishimoto2013ApJ...775L..36K, Kishimoto2022ApJ...940...28K}). We note that the total IR luminosity is mainly contributed by the host galaxy, as the nuclear star formation rate (SFR) is less than $0.06~\mathrm{ M_\odot~yr^{-1}}$ within the central 25 pc region \citep{Esquej2014ApJ...780...86E}, which is an order of magnitude lower than the total SFR \citep{Erroz-Ferrer2015MNRAS.451.1004E, Theios2016ApJ...822...45T}.
 
\section{Gamma-ray Emission From Jet and Corona}
For the jet emission, we consider synchrotron, synchrotron self-Compton (SSC), and external Compton (EC) from the C4 component. Synchrotron emission is calculated following the prescription of \citet{Aharonian2010PhRvD..82d3002A}, for SSC we use the differential cross section from \citet{1981Ap&SS..79..321A}, and the EC calculations are based on \citet{Khangulyan2014ApJ...783..100K}. We adopt the model of \citet{Mahmoud2020MNRAS.491.5126M} and the dust torus component \citep{Kishimoto2013ApJ...775L..36K} as the thermal disk component of NGC~4151, which serves as the target photon field of the EC. For the dust torus, we assume a blackbody spectrum with a temperature of $10^3$~K and normalize it to 133~mJy at the $K$-band. Since the jet velocity is non-relativistic and the jet viewing angle is large, the impact of IC scattering in the anisotropic regime (see, e.g., in \cite{Khangulyan2014ApJ...783..100K}) should be modest. Thus, we assume an isotropic photon field. 

We set a magnetic field strength of $B=4.2\times10^{-4}~\mathrm{G}$, which is approximately in concordance with the magnetic field estimate based on the minimum energy condition \citep{Williams2017MNRAS.472.3842W}. For the electron energy distribution, we assume a broken power-law spectral shape with a cutoff. The low-energy electron spectral index is set to $p_1=1.8$, given the spectral index of the C4 component \citep{Williams2017MNRAS.472.3842W}, we set it as $p_2=3.0$ at the high-energy end. The break Lorentz factor is set to $\gamma_\mathrm{br}=5.0\times10^4$, roughly consistent with the position of the synchrotron cooling break expected for the C4 dynamic time scale, $R/v_\mathrm{jet}$. We set the minimum and the maximum Lorentz factors as $\gamma_\mathrm{min}=1.0$ and $\gamma_\mathrm{cut}=4.0\times10^5$, respectively. We set the value of $\gamma_\mathrm{cut}$ to be consistent with the 30~GeV Fermi-LAT data point. We selected such a normalization factor for the electron distribution that the electron power matches the Poynting flux: $P_e=P_B=1.5\times10^{41}~\mathrm{erg~s^{-1}}$.

We adopt the model outlined by \citet{Inoue2019ApJ...880...40I} for the non-thermal coronal component. This model, developed in response to the discovery of non-thermal coronal synchrotron emission \citep{Inoue2018ApJ...869..114I}, successfully simulates non-thermal electron distributions, taking into account coronal properties identified by X-ray and ALMA observations. The model suggests a preferred injection spectral index of $2$, which aligns with expectations for diffusive shock acceleration \citep{Drury1983RPPh...46..973D, Blandford1987PhR...154....1B}. To match the ALMA flux and cosmic MeV gamma-ray background flux, the model sets the non-thermal lepton energy at $f_\mathrm{nth}=0.03$ of the thermal leptons. The hadronic component of the model assumes the same energy injection as in leptons, and then calculates $pp$ and $p\gamma$ interactions.

In this Letter, we incorporate the photon fields presented by \citet{Mahmoud2020MNRAS.491.5126M} and their associated physical scales. The X-ray spectral cutoff is around $150$keV with a continuum slope of approximately $1.7$. We use a corresponding coronal opacity of $\tau_X\simeq2.2$, based on the Kompaneets equation \citep{Zdziarski1996MNRAS.283..193Z}, and update the internal gamma-ray attenuation accordingly. The corona size is set to $45r_s$, where $r_s$ is the Schwarzschild radius, following \citet{Mahmoud2020MNRAS.491.5126M}. As for the coronal magnetic field strength, due to the lack of high spatial resolution millimeter observation for NGC~4151, it is challenging to determine the coronal magnetic field observationally using the coronal synchrotron emission method \citep{Inoue2014PASJ...66L...8I, Raginski2016MNRAS.459.2082R}. We, therefore, assume that the coronal magnetic field strength of NGC~4151 is similar to other nearby Seyferts, at $B=30~\mathrm{G}$ \citep{Inoue2018ApJ...869..114I, Michiyama2023arXiv230615950M}. We note that the estimated corona size in these Seyferts, based on ALMA data, is consistent with the size of NGC~4151 as derived by \citet{Mahmoud2020MNRAS.491.5126M}. Lastly, while keeping the injection particle spectral index $p_\mathrm{inj}=2.0$, we set the gyrofactor $\eta_g=2000$, which is a key parameter determining the maximum energy. This choice is made to avoid contradicting the Fermi-LAT data. We also note that based on a comparison with the IceCube data for NGC~1068 \citep{Inoue2020ApJ...891L..33I, IceCube2022Sci...378..538I}, a gyrofactor of around $\eta_g\sim1000$ is preferred for NGC~1068 to yield a soft neutrino spectrum at $\gtrsim$~TeV.

\begin{figure}[t]
\begin{center}
\includegraphics[width=\linewidth]{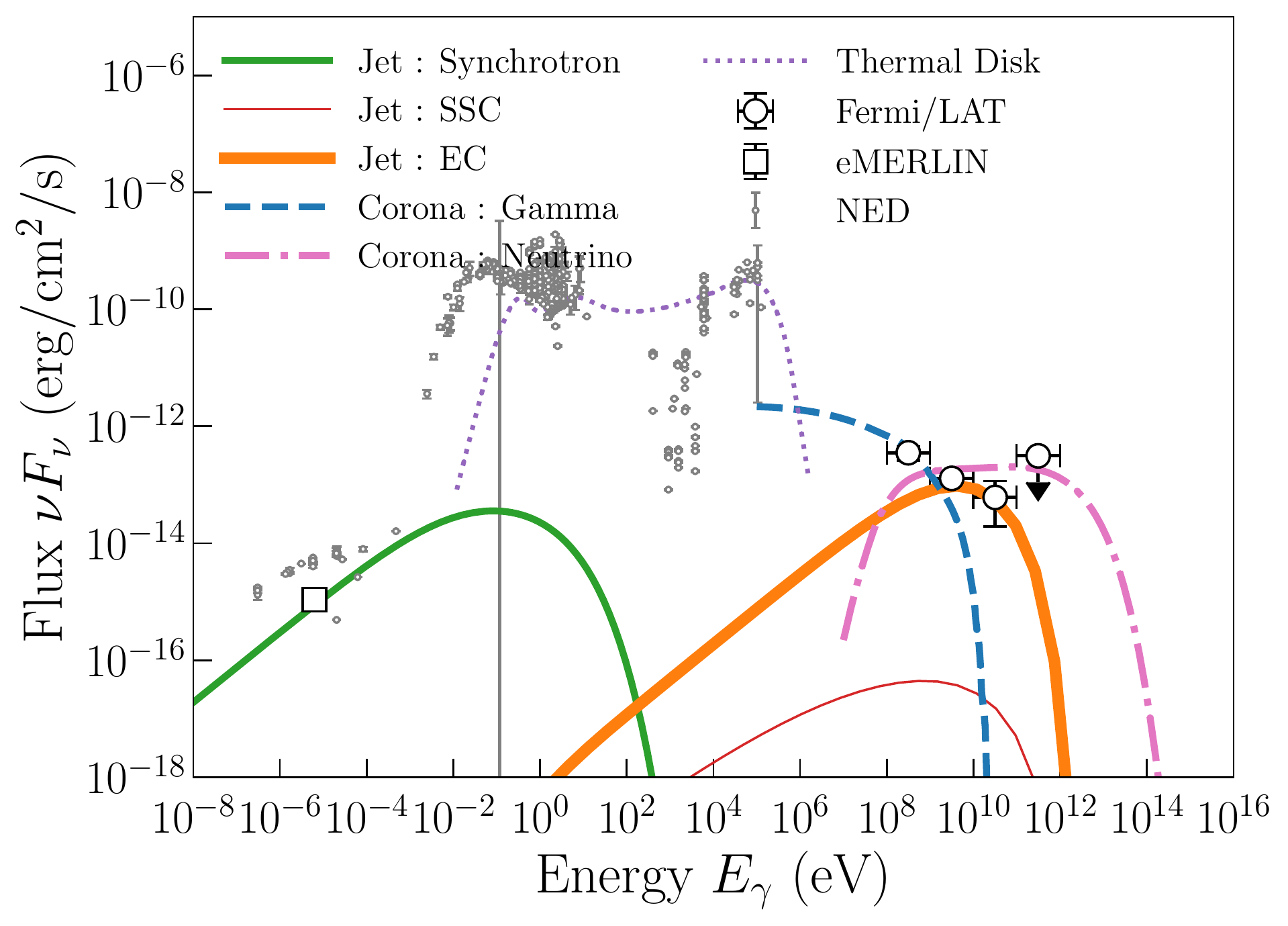}
\end{center} 
\caption{Multimessenger spectrum of NGC~4151.  The green, red, orange, blue, and magenta lines show the jet synchrotron, the jet SSC, the jet EC, the coronal gamma-ray, and the coronal neutrino spectrum, respectively. The purple line shows the thermal disk component that includes the Opt/UV/X-ray disk model of \citet{Mahmoud2020MNRAS.491.5126M} and the IR dust component \citep{Kishimoto2013ApJ...775L..36K}. 
 Square and circle symbols show the eMERLIN data \citep{Williams2017MNRAS.472.3842W} and the Fermi-LAT data \citep{Peretti2023arXiv230303298P}, respectively. Archival NED data are also shown in gray.
\label{fig:SED}}
\end{figure}

Fig.~\ref{fig:SED} shows the broadband model spectra accounting for jet synchrotron (green), jet SSC (red), jet EC (orange), coronal gamma ray (blue) and coronal neutrino (magenta). The eMERLIN data \citep{Williams2017MNRAS.472.3842W}, the Fermi-LAT data \citep{Peretti2023arXiv230303298P}, and the archival NED data are also presented. We also show the thermal disk component by \citet{Mahmoud2020MNRAS.491.5126M} together with the dust emission. Internal gamma-ray attenuation is taken into account in our modeling. Since the attenuation by the extragalactic background light is negligible at this distance of 15.8~Mpc, we do not include it here. As shown in the figure, the SSC component does not significantly contribute to the gamma-ray flux. We note that the non-thermal coronal model should smoothly connect to the thermal model in thermal and non-thermal hybrid models. However, simultaneous modeling of thermal and non-thermal components is beyond the scope of this work.

\section{Discussion}
\subsection{High-Energy Neutrino Flux Prediction}

\begin{figure}[t]
\begin{center}
\includegraphics[width=\linewidth]{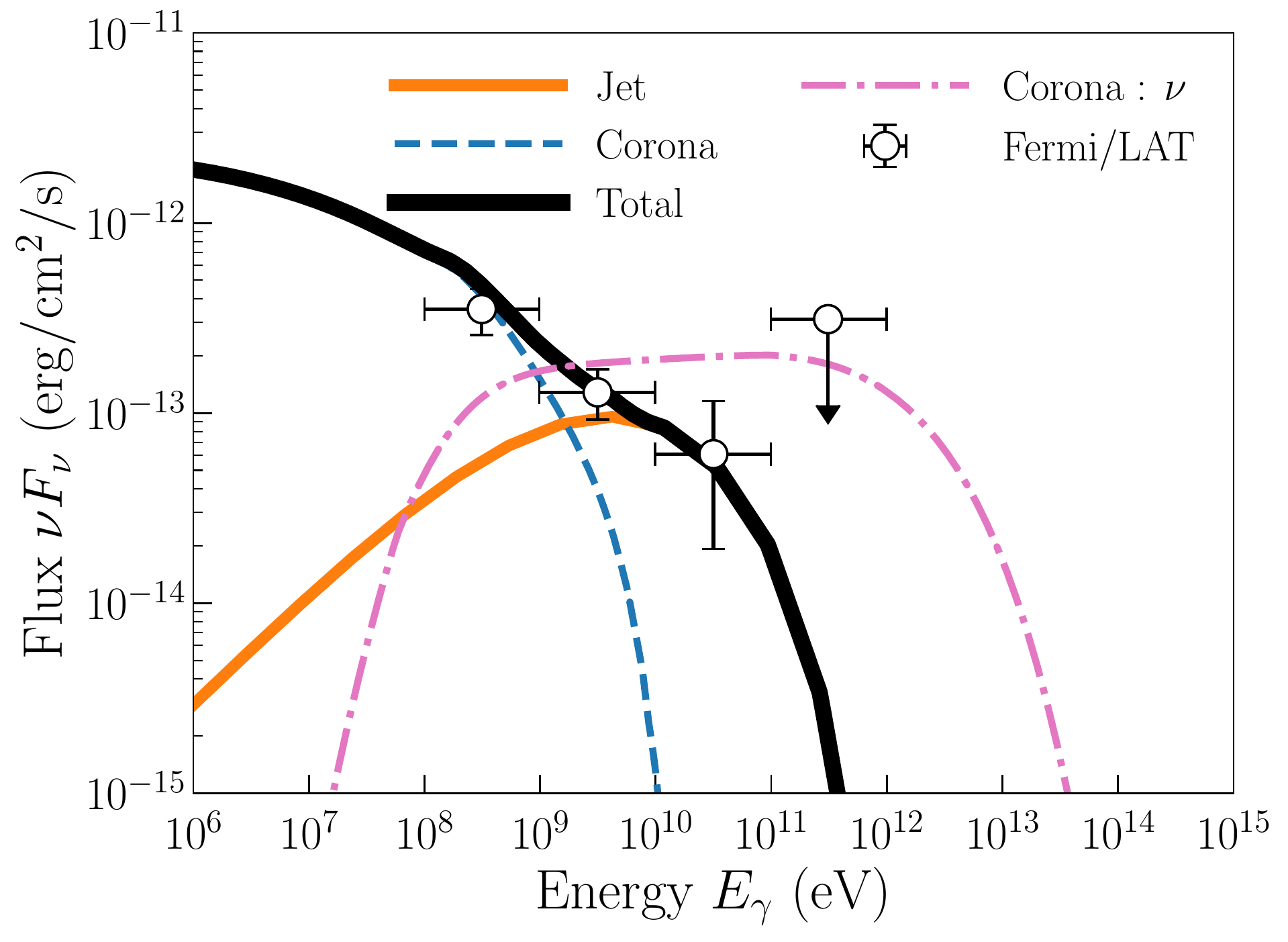}
\end{center} 
\caption{Same as Fig.~\ref{fig:SED} with $p_\mathrm{inj}=2.0$ and the gyrofactor $\eta_g = 2000$ for the coronal parameter, but showing the energy range only in $10^6$--$10^{15}$~eV together with the Fermi-LAT data \citep{Peretti2023arXiv230303298P}.
\label{fig:Zoom}}
\end{figure}

Fig.~\ref{fig:Zoom} provides a closer look at the energy range from $10^6$--$10^{15}$~eV, as depicted in Fig.~\ref{fig:SED}. The figure illustrates that the combined spectrum of the jet and corona aligns with the Fermi-LAT data. The anticipated neutrino flux above 10~TeV is approximately $10^{-13}~\mathrm{erg~cm^{-2}~s^{-1}}$. The $5\sigma$ sensitivity of IceCube towards NGC~4151 is $3.0\times10^{-11}~\mathrm{TeV^{-1}~cm^{-2}~s^{-1}}$ and $1.6\times10^{-12}~\mathrm{TeV^{-1}~cm^{-2}~s^{-1}}$ at 1~TeV for the $E^{-3}$ and $E^{-2}$ spectrum sources, respectively \citep{IceCube2020PhRvL.124e1103A}, which are still above the neutrino flux expected from the corona\footnote{After the submission of this paper, IceCube has reported a 2.9-$\sigma$ signal from the direction of NGC~4151 \citep{IceCube2023arXiv230715349G, IceCube2023arXiv230800024G}. Further detailed analysis and deeper observations with the next generation of neutrino telescopes are required to interpret the reported 2.9-$\sigma$ signal. Provided the limited angular resolution of neutrino telescopes, such a study should also include nearby objects, which are potential neutrino sources. In particular, Buson et al. (2023) reported that a blazar 5BZB J1210+3929 is located near NGC 4151.}. Based on our model, detecting neutrinos from the NGC~4151 corona would necessitate next-generation neutrino observatories like IceCube-Gen2 \citep{Clark+21} and KM3NeT \citep{Adrian-Martinez+16}. If IceCube reports neutrino signal from NGC~4151, it may imply the presence of other contributing factors, such as a harder particle injection spectrum expected for stochastic particle acceleration (see e.g., \cite{Murase2020PhRvL.125a1101M, Eichmann2022ApJ...939...43E}) or the presence of other contributing sources (like highly gamma-ray attenuated compact sources in NGC~4151 or other nearby extragalactic sources). 

\begin{figure}[t]
\begin{center}
\includegraphics[width=\linewidth]{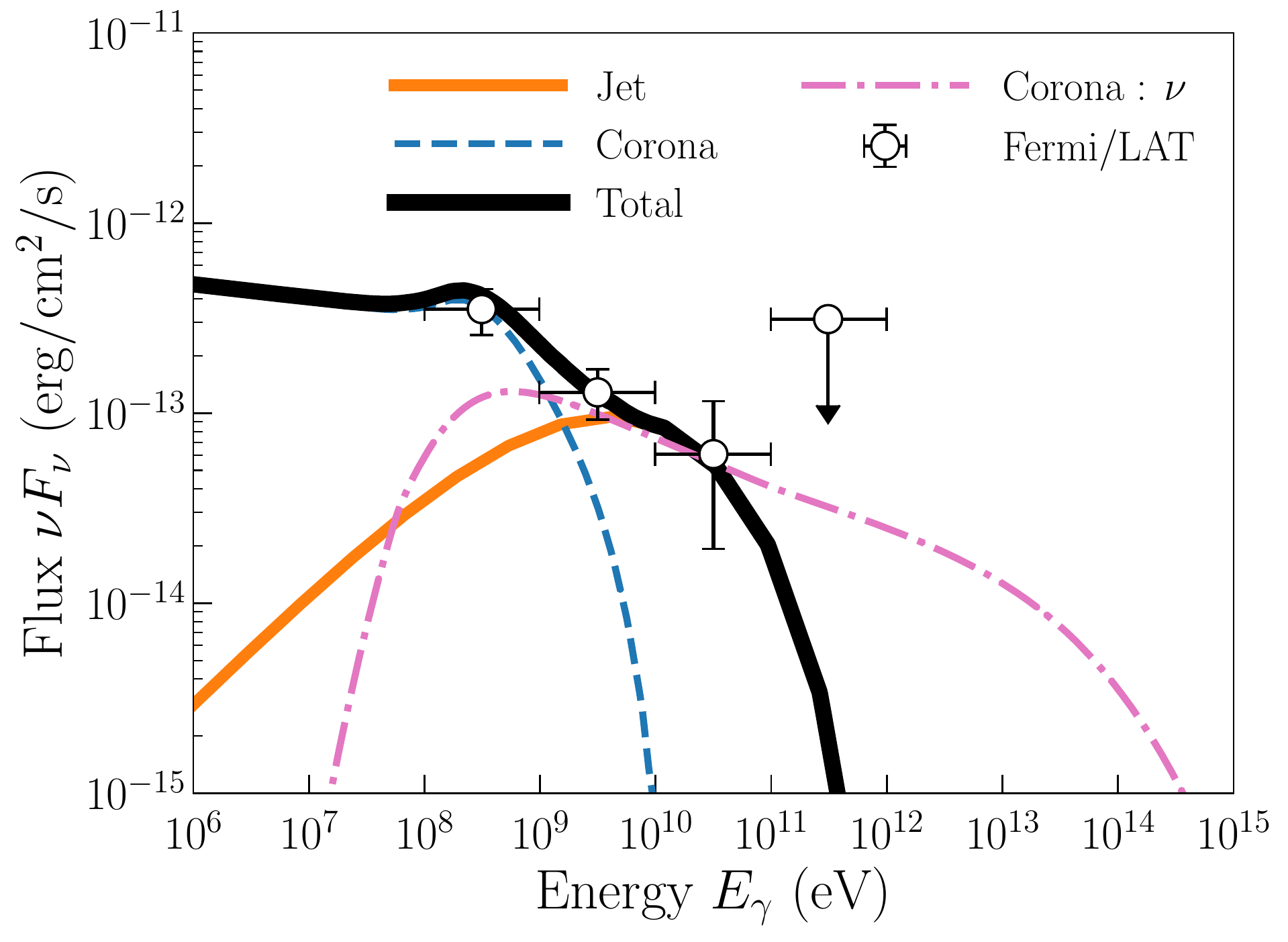}
\end{center} 
\caption{Same as Fig.~\ref{fig:Zoom}, but with $p_\mathrm{inj}=2.3$ and the gyrofactor $\eta_g = 30$.
\label{fig:Zoom_eta}}
\end{figure}

Fig.~\ref{fig:Zoom_eta} displays a plot similar to Fig.~\ref{fig:Zoom}, but with the injection particle spectral index set to $2.3$ and $\eta_g = 30$. This figure shows that  a low $\eta_g$ parameter, combined with a softer injection spectrum,  can also explain the data. In this case, the expected neutrino flux is significantly suppressed. Future MeV gamma-ray observations \citep{COSI2019BAAS...51g..98T, AMEGO2019BAAS...51g.245M, GECCO2022JCAP...07..036O, GRAMS2020APh...114..107A} could help resolve the ambiguity between $p_\mathrm{inj}$ and $\eta_g$.

Despite NGC~4151 being the X-ray brightest Seyfert, IceCube has already detected NGC~1068 at a 4.2$\sigma$ level. As discussed in \citet{Inoue2020ApJ...891L..33I}, this discrepancy can be attributed to the X-ray attenuation effect and the penetrating power of neutrinos. When we adjust for the X-ray attenuation effect, NGC~1068 emerges as the intrinsically brightest Seyfert, outshining NGC~4151 by a factor of approximately $3.6$. To align with the Fermi-LAT data, our model favors a high gyrofactor and/or a soft particle injection spectrum for NGC~4151, further diminishing the expected neutrino flux.

Certain numerical simulations predict stronger magnetic fields in the coronal region than our model assumes \citep{Liska2022ApJ...935L...1L}. These simulations often presuppose strong large-scale poloidal magnetic fields to reproduce the powerful jets seen in radio-loud AGNs successfully. However, NGC~4151 is a radio-quiet AGN, which lacks significant jet activity. To prevent the formation of powerful jets, a toroidal magnetic field configuration is preferred, resulting in weaker coronal magnetic fields than those associated with large-scale poloidal magnetic fields \citep{Liska2022ApJ...935L...1L}. Furthermore, in the presence of weak large-scale poloidal magnetic fields, the Parker instability \citep{Parker1955ApJ...121..491P, Parker1966ApJ...145..811P} could inhibit the amplification of the magnetic field \citep{Takasao2018ApJ...857....4T}.

\subsection{Jet Only Scenario}
\begin{figure}[t]
\begin{center}
\includegraphics[width=\linewidth]{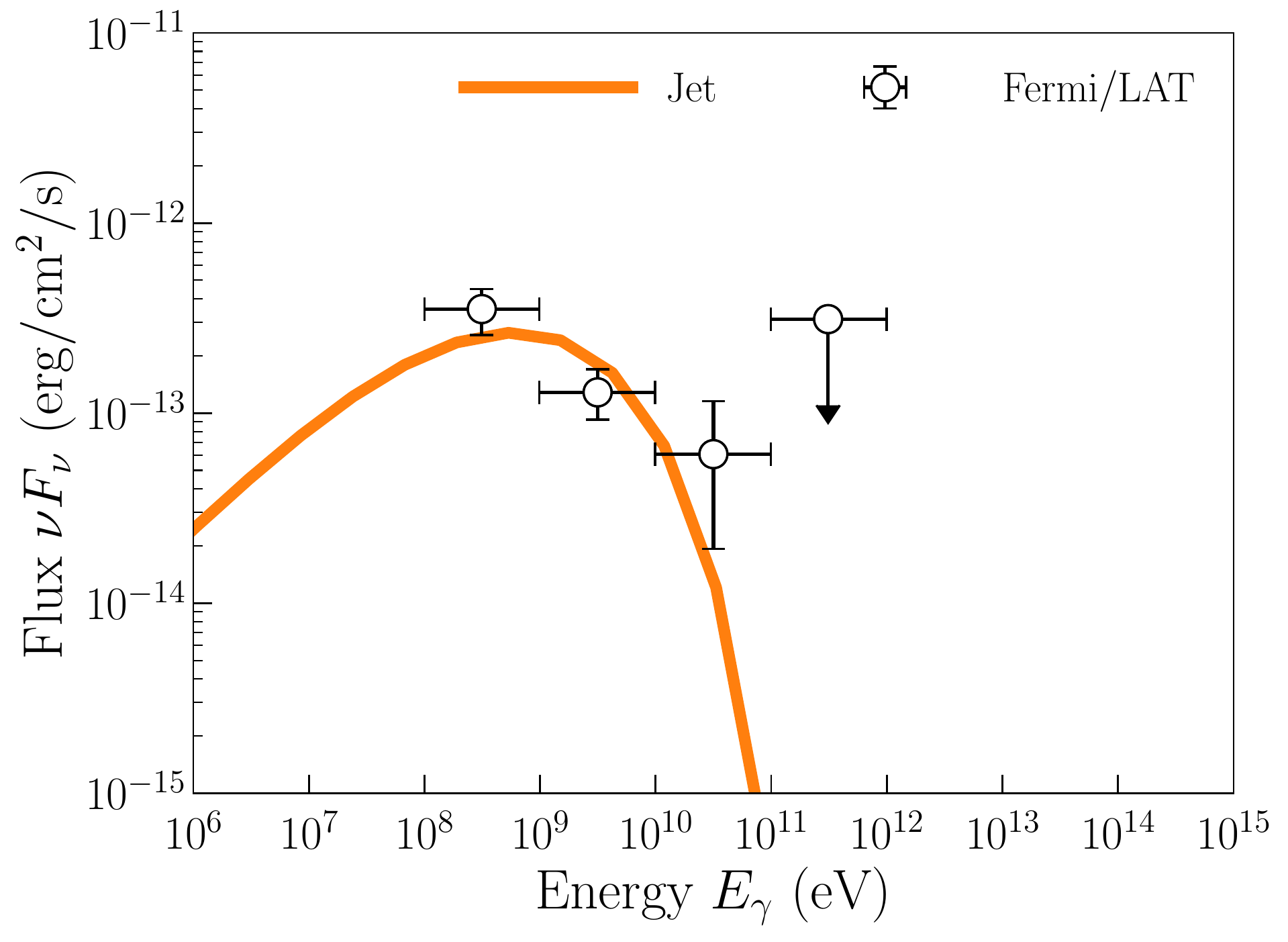}
\end{center} 
\caption{Same as Fig.~\ref{fig:Zoom}, but showing the jet alone model.
\label{fig:Zoom_jet}}
\end{figure}

The Fermi-LAT gamma-ray spectrum could be explained solely by the jet emission. Fig.~\ref{fig:Zoom_jet} shows the case only with the jet. We change the following jet parameters as follows: $B=1.2\times10^{-4}~\mathrm{G}$, $\gamma_\mathrm{br}=2.0\times10^4$, $\gamma_\mathrm{cut}=6.0\times10^4$, $P_B=1.2\times10^{40}~\mathrm{erg~s^{-1}}$ and $P_e=9.7\times10^{41}~\mathrm{erg~s^{-1}}$. We keep the other parameters as shown in Figs.~\ref{fig:SED} and \ref{fig:Zoom}. As compared to the corona+jet model, we require a matter-dominated jet $P_e/P_B\sim80$, where the magnetic field strength becomes a factor of $\sim10$ smaller than the minimum energy condition \citep{Williams2017MNRAS.472.3842W}. If such a matter-dominated jet is realized, non-thermal coronal activity should be suppressed.

Energy-dependent time variability would allow us to distinguish between the jet+corona and jet-only scenarios. The physical scale of the jet is $12$~pc, while that of the corona is $45~r_s$. As there is a $\sim10^5$ scale difference between these two regions, these two regions would produce different variability time scales. Since the corona can dominate only at $\lesssim1$~ GeV due to internal attenuation, the search for energy-dependent time variabilities would be the key to disentangling these two scenarios. More detailed time variability analysis by Fermi or future MeV gamma-ray missions \citep{COSI2019BAAS...51g..98T, AMEGO2019BAAS...51g.245M, GECCO2022JCAP...07..036O, GRAMS2020APh...114..107A} would be necessary.

Very Long Baseline Interferometry (VLBI) observations have revealed several components within the C4 component along the jet axis, with an angular resolution of a few milliarcseconds, corresponding to a physical scale of approximately 0.1~pc \citep{Ulvestad1998ApJ...496..196U, Ulvestad2005AJ....130..936U}. However, the reported flux densities of those components are only on the order of several mJy at 8~GHz. This suggests that diffuse emission spread over the C4 component would contribute more significantly to the integrated flux (72~mJy) measured by eMERLIN. Therefore, in this letter, we assume that the C4 component has a uniform emissivity.

\section{Summary}
Detection of gamma-ray emission from the X-ray brightest Seyfert NGC~4151 has been recently claimed by \citet{Peretti2023arXiv230303298P}, which is a puzzling discovery given the galaxy's low star formation rate. We propose that the emission could be produced by combining the coronal and jet activities of NGC~4151. Alternatively, jet activity alone may be able to explain the observed gamma-ray spectrum. To distinguish between these two scenarios, an energy-dependent variability search would be necessary, as the coronal component can only contribute at energies of $\lesssim1$ GeV because of the internal gamma-ray attenuation. Additionally, we analyze the expected neutrino flux from the coronal component, and our results suggest that it would likely be undetectable by IceCube, based on the Fermi-LAT spectrum.


\begin{ack}
We would like to thank the anonymous referee for thoughtful and helpful comments. 
We thank Yasushi Fukazawa, Hidetoshi Kubo, Tomonari Michiyama, Ellis Owen, and Shinsuke Takasao for useful discussions and comments. We also acknowledge the value of the "Topical Workshop: NGC 1068 as cosmic laboratory" sponsored by SFB1258 and Cluster of Excellence ORIGINS, which stimulated this work. YI is supported by JSPS KAKENHI Grant Numbers JP18H05458, JP19K14772, and JP22K18277. This work was supported by World Premier International Research Center Initiative (WPI), MEXT, Japan.  
\end{ack}


\begin{thebibliography}{72}
\expandafter\ifx\csname natexlab\endcsname\relax\def\natexlab#1{#1}\fi

\bibitem[{{Aartsen} {et~al.}(2020){Aartsen}, {Ackermann}, {Adams}, {Aguilar},
  {Ahlers}, {Ahrens}, {Alispach}, {Andeen}, {Anderson}, {Ansseau}, {Anton},
  {Arg{\"u}elles}, {Auffenberg}, {Axani}, {Backes}, {Bagherpour}, {Bai},
  {Balagopal}, {Barbano}, {Barwick}, {Bastian}, {Baum}, {Baur}, {Bay},
  {Beatty}, {Becker}, {Becker Tjus}, {BenZvi}, {Berley}, {Bernardini},
  {Besson}, {Binder}, {Bindig}, {Blaufuss}, {Blot}, {Bohm}, {B{\"o}rner},
  {B{\"o}ser}, {Botner}, {B{\"o}ttcher}, {Bourbeau}, {Bourbeau}, {Bradascio},
  {Braun}, {Bron}, {Brostean-Kaiser}, {Burgman}, {Buscher}, {Busse}, {Carver},
  {Chen}, {Cheung}, {Chirkin}, {Choi}, {Clark}, {Classen}, {Coleman}, {Collin},
  {Conrad}, {Coppin}, {Correa}, {Cowen}, {Cross}, {Dave}, {De Clercq},
  {DeLaunay}, {Dembinski}, {Deoskar}, {De Ridder}, {Desiati}, {de Vries}, {de
  Wasseige}, {de With}, {DeYoung}, {Diaz}, {D{\'\i}az-V{\'e}lez}, {Dujmovic},
  {Dunkman}, {Dvorak}, {Eberhardt}, {Ehrhardt}, {Eller}, {Engel}, {Evenson},
  {Fahey}, {Fazely}, {Felde}, {Filimonov}, {Finley}, {Fox}, {Franckowiak},
  {Friedman}, {Fritz}, {Gaisser}, {Gallagher}, {Ganster}, {Garrappa},
  {Gerhardt}, {Ghorbani}, {Glauch}, {Gl{\"u}senkamp}, {Goldschmidt},
  {Gonzalez}, {Grant}, {Griffith}, {Griswold}, {G{\"u}nder}, {G{\"u}nd{\"u}z},
  {Haack}, {Hallgren}, {Halliday}, {Halve}, {Halzen}, {Hanson}, {Haungs},
  {Hebecker}, {Heereman}, {Heix}, {Helbing}, {Hellauer}, {Henningsen},
  {Hickford}, {Hignight}, {Hill}, {Hoffman}, {Hoffmann}, {Hoinka},
  {Hokanson-Fasig}, {Hoshina}, {Huang}, {Huber}, {Huber}, {Hultqvist},
  {H{\"u}nnefeld}, {Hussain}, {In}, {Iovine}, {Ishihara}, {Japaridze}, {Jeong},
  {Jero}, {Jones}, {Jonske}, {Joppe}, {Kang}, {Kang}, {Kappes}, {Kappesser},
  {Karg}, {Karl}, {Karle}, {Katz}, {Kauer}, {Kelley}, {Kheirandish}, {Kim},
  {Kintscher}, {Kiryluk}, {Kittler}, {Klein}, {Koirala}, {Kolanoski},
  {K{\"o}pke}, {Kopper}, {Kopper}, {Koskinen}, {Kowalski}, {Krings},
  {Kr{\"u}ckl}, {Kulacz}, {Kurahashi}, {Kyriacou}, {Labare}, {Lanfranchi},
  {Larson}, {Lauber}, {Lazar}, {Leonard}, {Leszczy{\'n}ska}, {Leuermann},
  {Liu}, {Lohfink}, {Lozano Mariscal}, {Lu}, {Lucarelli}, {L{\"u}nemann},
  {Luszczak}, {Lyu}, {Ma}, {Madsen}, {Maggi}, {Mahn}, {Makino}, {Mallik},
  {Mallot}, {Mancina}, {Mari{\c{s}}}, {Maruyama}, {Mase}, {Matis}, {Maunu},
  {McNally}, {Meagher}, {Medici}, {Medina}, {Meier}, {Meighen-Berger}, {Menne},
  {Merino}, {Meures}, {Micallef}, {Mockler}, {Moment{\'e}}, {Montaruli},
  {Moore}, {Morse}, {Moulai}, {Muth}, {Nagai}, {Naumann}, {Neer},
  {Niederhausen}, {Nisa}, {Nowicki}, {Nygren}, {Obertacke Pollmann}, {Oehler},
  {Olivas}, {O'Murchadha}, {O'Sullivan}, {Palczewski}, {Pandya}, {Pankova},
  {Park}, {Peiffer}, {P{\'e}rez de los Heros}, {Philippen}, {Pieloth}, {Pinat},
  {Pizzuto}, {Plum}, {Porcelli}, {Price}, {Przybylski}, {Raab}, {Raissi},
  {Rameez}, {Rauch}, {Rawlins}, {Rea}, {Reimann}, {Relethford}, {Renschler},
  {Renzi}, {Resconi}, {Rhode}, {Richman}, {Robertson}, {Rongen}, {Rott},
  {Ruhe}, {Ryckbosch}, {Rysewyk}, {Safa}, {Sanchez Herrera}, {Sandrock},
  {Sandroos}, {Santander}, {Sarkar}, {Sarkar}, {Satalecka}, {Schaufel},
  {Schieler}, {Schlunder}, {Schmidt}, {Schneider}, {Schneider}, {Schr{\"o}der},
  {Schumacher}, {Sclafani}, {Seckel}, {Seunarine}, {Shefali}, {Silva},
  {Snihur}, {Soedingrekso}, {Soldin}, {Song}, {Spiczak}, {Spiering},
  {Stachurska}, {Stamatikos}, {Stanev}, {Stein}, {Steinm{\"u}ller}, {Stettner},
  {Steuer}, {Stezelberger}, {Stokstad}, {St{\"o}{\ss}l}, {Strotjohann},
  {St{\"u}rwald}, {Stuttard}, {Sullivan}, {Taboada}, {Tenholt}, {Ter-Antonyan},
  {Terliuk}, {Tilav}, {Tollefson}, {Tomankova}, {T{\"o}nnis}, {Toscano},
  {Tosi}, {Trettin}, {Tselengidou}, {Tung}, {Turcati}, {Turcotte}, {Turley},
  {Ty}, {Unger}, {Unland Elorrieta}, {Usner}, {Vandenbroucke}, {Van Driessche},
  {van Eijk}, {van Eijndhoven}, {Vanheule}, {van Santen}, {Vraeghe}, {Walck},
  {Wallace}, {Wallraff}, {Wandkowsky}, {Watson}, {Weaver}, {Weindl}, {Weiss},
  {Weldert}, {Wendt}, {Werthebach}, {Whelan}, {Whitehorn}, {Wiebe}, {Wiebusch},
  {Wille}, {Williams}, {Wills}, {Wolf}, {Wood}, {Wood}, {Woschnagg}, {Wrede},
  {Xu}, {Xu}, {Xu}, {Yanez}, {Yodh}, {Yoshida}, {Yuan}, \&
  {Z{\"o}cklein}}]{IceCube2020PhRvL.124e1103A}
{Aartsen}, M.~G., {Ackermann}, M., {Adams}, J., {et~al.} 2020, \prl, 124,
  051103

\bibitem[{{Abdollahi} {et~al.}(2022){Abdollahi}, {Acero}, {Baldini}, {Ballet},
  {Bastieri}, {Bellazzini}, {Berenji}, {Berretta}, {Bissaldi}, {Blandford},
  {Bloom}, {Bonino}, {Brill}, {Britto}, {Bruel}, {Burnett}, {Buson}, {Cameron},
  {Caputo}, {Caraveo}, {Castro}, {Chaty}, {Cheung}, {Chiaro}, {Cibrario},
  {Ciprini}, {Coronado-Bl{\'a}zquez}, {Crnogorcevic}, {Cutini}, {D'Ammando},
  {De Gaetano}, {Digel}, {Di Lalla}, {Dirirsa}, {Di Venere}, {Dom{\'\i}nguez},
  {Fallah Ramazani}, {Fegan}, {Ferrara}, {Fiori}, {Fleischhack}, {Franckowiak},
  {Fukazawa}, {Funk}, {Fusco}, {Galanti}, {Gammaldi}, {Gargano}, {Garrappa},
  {Gasparrini}, {Giacchino}, {Giglietto}, {Giordano}, {Giroletti}, {Glanzman},
  {Green}, {Grenier}, {Grondin}, {Guillemot}, {Guiriec}, {Gustafsson},
  {Harding}, {Hays}, {Hewitt}, {Horan}, {Hou}, {J{\'o}hannesson}, {Karwin},
  {Kayanoki}, {Kerr}, {Kuss}, {Landriu}, {Larsson}, {Latronico},
  {Lemoine-Goumard}, {Li}, {Liodakis}, {Longo}, {Loparco}, {Lott}, {Lubrano},
  {Maldera}, {Malyshev}, {Manfreda}, {Mart{\'\i}-Devesa}, {Mazziotta}, {Mereu},
  {Meyer}, {Michelson}, {Mirabal}, {Mitthumsiri}, {Mizuno}, {Moiseev},
  {Monzani}, {Morselli}, {Moskalenko}, {Negro}, {Nuss}, {Omodei}, {Orienti},
  {Orlando}, {Paneque}, {Pei}, {Perkins}, {Persic}, {Pesce-Rollins},
  {Petrosian}, {Pillera}, {Poon}, {Porter}, {Principe}, {Rain{\`o}}, {Rando},
  {Rani}, {Razzano}, {Razzaque}, {Reimer}, {Reimer}, {Reposeur},
  {S{\'a}nchez-Conde}, {Saz Parkinson}, {Scotton}, {Serini}, {Sgr{\`o}},
  {Siskind}, {Smith}, {Spandre}, {Spinelli}, {Sueoka}, {Suson}, {Tajima},
  {Tak}, {Thayer}, {Thompson}, {Torres}, {Troja}, {Valverde}, {Wood}, \&
  {Zaharijas}}]{Fermi2022ApJS..260...53A}
{Abdollahi}, S., {Acero}, F., {Baldini}, L., {et~al.} 2022, \apjs, 260, 53

\bibitem[{{Ackermann} {et~al.}(2012{\natexlab{a}}){Ackermann}, {Ajello},
  {Allafort}, {Baldini}, {Ballet}, {Barbiellini}, {Bastieri}, {Bechtol},
  {Bellazzini}, {Berenji}, {Bloom}, {Bonamente}, {Borgland}, {Bregeon},
  {Brigida}, {Bruel}, {Buehler}, {Buson}, {Caliandro}, {Cameron}, {Caraveo},
  {Casandjian}, {Cavazzuti}, {Cecchi}, {Charles}, {Chekhtman}, {Cheung},
  {Chiang}, {Ciprini}, {Claus}, {Cohen-Tanugi}, {Conrad}, {Cutini},
  {D'Ammando}, {de Angelis}, {de Palma}, {Dermer}, {Silva}, {Drell},
  {Drlica-Wagner}, {Enoto}, {Favuzzi}, {Fegan}, {Ferrara}, {Fortin},
  {Fukazawa}, {Fusco}, {Gargano}, {Gasparrini}, {Gehrels}, {Germani},
  {Giglietto}, {Giommi}, {Giordano}, {Giroletti}, {Godfrey}, {Grove},
  {Guiriec}, {Hadasch}, {Hayashida}, {Hays}, {Hughes}, {J{\'o}hannesson},
  {Johnson}, {Kamae}, {Katagiri}, {Kataoka}, {Kn{\"o}dlseder}, {Kuss}, {Lande},
  {Llena Garde}, {Longo}, {Loparco}, {Lott}, {Lovellette}, {Lubrano},
  {Madejski}, {Mazziotta}, {Michelson}, {Mizuno}, {Monte}, {Monzani},
  {Morselli}, {Moskalenko}, {Murgia}, {Nishino}, {Norris}, {Nuss}, {Ohno},
  {Ohsugi}, {Okumura}, {Orlando}, {Ozaki}, {Paneque}, {Pesce-Rollins},
  {Pierbattista}, {Piron}, {Pivato}, {Porter}, {Rain{\`o}}, {Rando}, {Razzano},
  {Reimer}, {Reimer}, {Ritz}, {Roth}, {Sanchez}, {Sbarra}, {Sgr{\`o}},
  {Siskind}, {Spandre}, {Spinelli}, {Stawarz}, {Strong}, {Takahashi},
  {Takahashi}, {Tanaka}, {Thayer}, {Thompson}, {Tibaldo}, {Tinivella},
  {Torres}, {Tosti}, {Troja}, {Uchiyama}, {Usher}, {Vandenbroucke},
  {Vasileiou}, {Vianello}, {Vitale}, {Waite}, {Winer}, {Wood}, {Wood}, {Yang},
  \& {Zimmer}}]{Fermi2012ApJ...747..104A}
{Ackermann}, M., {Ajello}, M., {Allafort}, A., {et~al.} 2012{\natexlab{a}},
  \apj, 747, 104

\bibitem[{{Ackermann} {et~al.}(2012{\natexlab{b}}){Ackermann}, {Ajello},
  {Allafort}, {Baldini}, {Ballet}, {Bastieri}, {Bechtol}, {Bellazzini},
  {Berenji}, {Bloom}, {Bonamente}, {Borgland}, {Bouvier}, {Bregeon}, {Brigida},
  {Bruel}, {Buehler}, {Buson}, {Caliandro}, {Cameron}, {Caraveo}, {Casandjian},
  {Cecchi}, {Charles}, {Chekhtman}, {Cheung}, {Chiang}, {Cillis}, {Ciprini},
  {Claus}, {Cohen-Tanugi}, {Conrad}, {Cutini}, {de Palma}, {Dermer}, {Digel},
  {Silva}, {Drell}, {Drlica-Wagner}, {Favuzzi}, {Fegan}, {Fortin}, {Fukazawa},
  {Funk}, {Fusco}, {Gargano}, {Gasparrini}, {Germani}, {Giglietto}, {Giordano},
  {Glanzman}, {Godfrey}, {Grenier}, {Guiriec}, {Gustafsson}, {Hadasch},
  {Hayashida}, {Hays}, {Hughes}, {J{\'o}hannesson}, {Johnson}, {Kamae},
  {Katagiri}, {Kataoka}, {Kn{\"o}dlseder}, {Kuss}, {Lande}, {Longo}, {Loparco},
  {Lott}, {Lovellette}, {Lubrano}, {Madejski}, {Martin}, {Mazziotta},
  {McEnery}, {Michelson}, {Mizuno}, {Monte}, {Monzani}, {Morselli},
  {Moskalenko}, {Murgia}, {Nishino}, {Norris}, {Nuss}, {Ohno}, {Ohsugi},
  {Okumura}, {Omodei}, {Orlando}, {Ozaki}, {Parent}, {Persic}, {Pesce-Rollins},
  {Petrosian}, {Pierbattista}, {Piron}, {Pivato}, {Porter}, {Rain{\`o}},
  {Rando}, {Razzano}, {Reimer}, {Reimer}, {Ritz}, {Roth}, {Sbarra}, {Sgr{\`o}},
  {Siskind}, {Spandre}, {Spinelli}, {Stawarz}, {Strong}, {Takahashi}, {Tanaka},
  {Thayer}, {Tibaldo}, {Tinivella}, {Torres}, {Tosti}, {Troja}, {Uchiyama},
  {Vandenbroucke}, {Vianello}, {Vitale}, {Waite}, {Wood}, \&
  {Yang}}]{Fermi2012ApJ...755..164A}
{Ackermann}, M., {Ajello}, M., {Allafort}, A., {et~al.} 2012{\natexlab{b}},
  \apj, 755, 164

\bibitem[{{Adri{\'a}n-Mart{\'\i}nez} {et~al.}(2016){Adri{\'a}n-Mart{\'\i}nez},
  {Ageron}, {Aharonian}, {Aiello}, {Albert}, {Ameli}, {Anassontzis}, {Andre},
  {Androulakis}, {Anghinolfi}, {Anton}, {Ardid}, {Avgitas}, {Barbarino},
  {Barbarito}, {Baret}, {Barrios-Mart{\'\i}}, {Belhorma}, {Belias}, {Berbee},
  {van den Berg}, {Bertin}, {Beurthey}, {van Beveren}, {Beverini}, {Biagi},
  {Biagioni}, {Billault}, {Bond{\`\i}}, {Bormuth}, {Bouhadef}, {Bourlis},
  {Bourret}, {Boutonnet}, {Bouwhuis}, {Bozza}, {Bruijn}, {Brunner}, {Buis},
  {Busto}, {Cacopardo}, {Caillat}, {Calamai}, {Calvo}, {Capone}, {Caramete},
  {Cecchini}, {Celli}, {Champion}, {Cherkaoui El Moursli}, {Cherubini},
  {Chiarusi}, {Circella}, {Classen}, {Cocimano}, {Coelho}, {Coleiro},
  {Colonges}, {Coniglione}, {Cordelli}, {Cosquer}, {Coyle}, {Creusot},
  {Cuttone}, {D'Amico}, {De Bonis}, {De Rosa}, {De Sio}, {Di Capua}, {Di
  Palma}, {D{\'\i}az Garc{\'\i}a}, {Distefano}, {Donzaud}, {Dornic},
  {Dorosti-Hasankiadeh}, {Drakopoulou}, {Drouhin}, {Drury}, {Durocher},
  {Eberl}, {Eichie}, {van Eijk}, {El Bojaddaini}, {El Khayati}, {Elsaesser},
  {Enzenh{\"o}fer}, {Fassi}, {Favali}, {Fermani}, {Ferrara}, {Filippidis},
  {Frascadore}, {Fusco}, {Gal}, {Galat{\`a}}, {Garufi}, {Gay}, {Gebyehu},
  {Giordano}, {Gizani}, {Gracia}, {Graf}, {Gr{\'e}goire}, {Grella}, {Habel},
  {Hallmann}, {van Haren}, {Harissopulos}, {Heid}, {Heijboer}, {Heine},
  {Henry}, {Hern{\'a}ndez-Rey}, {Hevinga}, {Hofest{\"a}dt}, {Hugon},
  {Illuminati}, {James}, {Jansweijer}, {Jongen}, {de Jong}, {Kadler},
  {Kalekin}, {Kappes}, {Katz}, {Keller}, {Kieft}, {Kie{\ss}ling}, {Koffeman},
  {Kooijman}, {Kouchner}, {Kulikovskiy}, {Lahmann}, {Lamare}, {Leisos},
  {Leonora}, {Clark}, {Liolios}, {Llorens Alvarez}, {Lo Presti}, {L{\"o}hner},
  {Lonardo}, {Lotze}, {Loucatos}, {Maccioni}, {Mannheim}, {Margiotta},
  {Marinelli}, {Mari{\c{s}}}, {Markou}, {Mart{\'\i}nez-Mora}, {Martini},
  {Mele}, {Melis}, {Michael}, {Migliozzi}, {Migneco}, {Mijakowski}, {Miraglia},
  {Mollo}, {Mongelli}, {Morganti}, {Moussa}, {Musico}, {Musumeci}, {Navas},
  {Nicolau}, {Olcina}, {Olivetto}, {Orlando}, {Papaikonomou}, {Papaleo},
  {P{\u{a}}v{\u{a}}la{\c{s}}}, {Peek}, {Pellegrino}, {Perrina}, {Pfutzner},
  {Piattelli}, {Pikounis}, {Poma}, {Popa}, {Pradier}, {Pratolongo},
  {P{\"u}hlhofer}, {Pulvirenti}, {Quinn}, {Racca}, {Raffaelli}, {Randazzo},
  {Rapidis}, {Razis}, {Real}, {Resvanis}, {Reubelt}, {Riccobene}, {Rossi},
  {Rovelli}, {Salda{\~n}a}, {Salvadori}, {Samtleben}, {S{\'a}nchez
  Garc{\'\i}a}, {S{\'a}nchez Losa}, {Sanguineti}, {Santangelo}, {Santonocito},
  {Sapienza}, {Schimmel}, {Schmelling}, {Sciacca}, {Sedita}, {Seitz}, {Sgura},
  {Simeone}, {Siotis}, {Sipala}, {Spisso}, {Spurio}, {Stavropoulos},
  {Steijger}, {Stellacci}, {Stransky}, {Taiuti}, {Tayalati}, {T{\'e}zier},
  {Theraube}, {Thompson}, {Timmer}, {T{\"o}nnis}, {Trasatti}, {Trovato},
  {Tsirigotis}, {Tzamarias}, {Tzamariudaki}, {Vallage}, {Van Elewyck},
  {Vermeulen}, {Vicini}, {Viola}, {Vivolo}, {Volkert}, {Voulgaris}, {Wiggers},
  {Wilms}, {de Wolf}, {Zachariadou}, {Zornoza}, \&
  {Z{\'u}{\~n}iga}}]{Adrian-Martinez+16}
{Adri{\'a}n-Mart{\'\i}nez}, S., {Ageron}, M., {Aharonian}, F., {et~al.} 2016,
  Journal of Physics G Nuclear Physics, 43, 084001

\bibitem[{{Aharonian} \& {Atoyan}(1981)}]{1981Ap&SS..79..321A}
{Aharonian}, F.~A. \& {Atoyan}, A.~M. 1981, \apss, 79, 321

\bibitem[{{Aharonian} {et~al.}(2010){Aharonian}, {Kelner}, \&
  {Prosekin}}]{Aharonian2010PhRvD..82d3002A}
{Aharonian}, F.~A., {Kelner}, S.~R., \& {Prosekin}, A.~Y. 2010, \prd, 82,
  043002

\bibitem[{{Ajello} {et~al.}(2021){Ajello}, {Baldini}, {Ballet}, {Barbiellini},
  {Bastieri}, {Bellazzini}, {Berretta}, {Bissaldi}, {Blandford}, {Bloom},
  {Bonino}, {Bruel}, {Buson}, {Cameron}, {Caprioli}, {Caputo}, {Cavazzuti},
  {Chartas}, {Chen}, {Cheung}, {Chiaro}, {Costantin}, {Cutini}, {D'Ammando},
  {de la Torre Luque}, {de Palma}, {Desai}, {Diesing}, {Di Lalla}, {Dirirsa},
  {Di Venere}, {Dom{\'\i}nguez}, {Fegan}, {Franckowiak}, {Fukazawa}, {Funk},
  {Fusco}, {Gargano}, {Gasparrini}, {Giglietto}, {Giordano}, {Giroletti},
  {Green}, {Grenier}, {Guiriec}, {Hartmann}, {Horan}, {J{\'o}hannesson},
  {Karwin}, {Kerr}, {Kova{\v{c}}evi{\'c}}, {Kuss}, {Larsson}, {Latronico},
  {Lemoine-Goumard}, {Li}, {Liodakis}, {Longo}, {Loparco}, {Lovellette},
  {Lubrano}, {Maldera}, {Manfreda}, {Marchesi}, {Marcotulli},
  {Mart{\'\i}-Devesa}, {Mazziotta}, {Mereu}, {Michelson}, {Mizuno}, {Monzani},
  {Morselli}, {Moskalenko}, {Negro}, {Omodei}, {Orienti}, {Orlando}, {Paliya},
  {Paneque}, {Pei}, {Persic}, {Pesce-Rollins}, {Porter}, {Principe}, {Racusin},
  {Rain{\`o}}, {Rando}, {Rani}, {Razzano}, {Reimer}, {Reimer}, {Saz Parkinson},
  {Serini}, {Sgr{\`o}}, {Siskind}, {Spandre}, {Spinelli}, {Suson}, {Tak},
  {Torres}, {Troja}, {Wood}, {Zaharijas}, \&
  {Zrake}}]{Ajello2021ApJ...921..144A}
{Ajello}, M., {Baldini}, L., {Ballet}, J., {et~al.} 2021, \apj, 921, 144

\bibitem[{{Ajello} {et~al.}(2020){Ajello}, {Di Mauro}, {Paliya}, \&
  {Garrappa}}]{Ajello2020ApJ...894...88A}
{Ajello}, M., {Di Mauro}, M., {Paliya}, V.~S., \& {Garrappa}, S. 2020, \apj,
  894, 88

\bibitem[{{Aramaki} {et~al.}(2020){Aramaki}, {Adrian}, {Karagiorgi}, \&
  {Odaka}}]{GRAMS2020APh...114..107A}
{Aramaki}, T., {Adrian}, P. O.~H., {Karagiorgi}, G., \& {Odaka}, H. 2020,
  Astroparticle Physics, 114, 107

\bibitem[{{Bignami} {et~al.}(1979){Bignami}, {Fichtel}, {Hartman}, \&
  {Thompson}}]{Bignami1979ApJ...232..649B}
{Bignami}, G.~F., {Fichtel}, C.~E., {Hartman}, R.~C., \& {Thompson}, D.~J.
  1979, \apj, 232, 649

\bibitem[{{Blandford} \& {Eichler}(1987)}]{Blandford1987PhR...154....1B}
{Blandford}, R. \& {Eichler}, D. 1987, \physrep, 154, 1

\bibitem[{{Carral} {et~al.}(1990){Carral}, {Turner}, \&
  {Ho}}]{Carral1990ApJ...362..434C}
{Carral}, P., {Turner}, J.~L., \& {Ho}, P. T.~P. 1990, \apj, 362, 434

\bibitem[{{Clark} {et~al.}(2021){Clark}, {Clark}, \& {IceCube-Gen2
  Collaboration}}]{Clark+21}
{Clark}, B.~A., {Clark}, B.~A., \& {IceCube-Gen2 Collaboration}. 2021, Journal
  of Instrumentation, 16, C10007

\bibitem[{{Couto} {et~al.}(2016){Couto}, {Kraemer}, {Turner}, \&
  {Crenshaw}}]{Couto2016ApJ...833..191C}
{Couto}, J.~D., {Kraemer}, S.~B., {Turner}, T.~J., \& {Crenshaw}, D.~M. 2016,
  \apj, 833, 191

\bibitem[{{De Rosa} {et~al.}(2018){De Rosa}, {Fausnaugh}, {Grier}, {Peterson},
  {Denney}, {Horne}, {Bentz}, {Ciroi}, {Dalla Bont{\`a}}, {Joner}, {Kaspi},
  {Kochanek}, {Pogge}, {Sergeev}, {Vestergaard}, {Adams}, {Antognini}, {Araya
  Salvo}, {Armstrong}, {Bae}, {Barth}, {Beatty}, {Bhattacharjee}, {Borman},
  {Boroson}, {Bottorff}, {Brown}, {Brown}, {Brotherton}, {Coker}, {Clanton},
  {Cracco}, {Crawford}, {Croxall}, {Eftekharzadeh}, {Eracleous}, {Fiorenza},
  {Frassati}, {Hawkins}, {Henderson}, {Holoien}, {Hutchison}, {Kellar},
  {Kilerci-Eser}, {Kim}, {King}, {La Mura}, {Laney}, {Li}, {Lochhaas}, {Ma},
  {MacInnis}, {Manne-Nicholas}, {Mason}, {McGraw}, {Mogren}, {Montouri},
  {Moody}, {Mosquera}, {Mudd}, {Musso}, {Nazarov}, {Nguyen}, {Ochner},
  {Okhmat}, {Onken}, {Ou-Yang}, {Pancoast}, {Pei}, {Penny}, {Poleski},
  {Portaluri}, {Prieto}, {Price-Whelan}, {Pulatova}, {Rafter}, {Roettenbacher},
  {Romero-Colmenero}, {Runnoe}, {Schimoia}, {Shappee}, {Sherf}, {Simonian},
  {Siviero}, {Skowron}, {Skowron}, {Somers}, {Spencer}, {Starkey}, {Stevens},
  {Stoll}, {Tamajo}, {Tayar}, {van Saders}, {Valenti}, {Villanueva},
  {Villforth}, {Weiss}, {Winkler}, {Zastrow}, {Zhu}, \&
  {Zu}}]{DeRosa2018ApJ...866..133D}
{De Rosa}, G., {Fausnaugh}, M.~M., {Grier}, C.~J., {et~al.} 2018, \apj, 866,
  133

\bibitem[{{de Vaucouleurs} {et~al.}(1981){de Vaucouleurs}, {Peters},
  {Bottinelli}, {Gouguenheim}, \& {Paturel}}]{deVaucouleurs1981ApJ...248..408D}
{de Vaucouleurs}, G., {Peters}, W.~L., {Bottinelli}, L., {Gouguenheim}, L., \&
  {Paturel}, G. 1981, \apj, 248, 408

\bibitem[{{Drury}(1983)}]{Drury1983RPPh...46..973D}
{Drury}, L.~O. 1983, Reports on Progress in Physics, 46, 973

\bibitem[{{Edelson} {et~al.}(2017){Edelson}, {Gelbord}, {Cackett}, {Connolly},
  {Done}, {Fausnaugh}, {Gardner}, {Gehrels}, {Goad}, {Horne}, {McHardy},
  {Peterson}, {Vaughan}, {Vestergaard}, {Breeveld}, {Barth}, {Bentz},
  {Bottorff}, {Brandt}, {Crawford}, {Dalla Bont{\`a}}, {Emmanoulopoulos},
  {Evans}, {Figuera Jaimes}, {Filippenko}, {Ferland}, {Grupe}, {Joner},
  {Kennea}, {Korista}, {Krimm}, {Kriss}, {Leonard}, {Mathur}, {Netzer},
  {Nousek}, {Page}, {Romero-Colmenero}, {Siegel}, {Starkey}, {Treu}, {Vogler},
  {Winkler}, \& {Zheng}}]{Edelson2017ApJ...840...41E}
{Edelson}, R., {Gelbord}, J., {Cackett}, E., {et~al.} 2017, \apj, 840, 41

\bibitem[{{Eichmann} {et~al.}(2022){Eichmann}, {Oikonomou}, {Salvatore},
  {Dettmar}, \& {Tjus}}]{Eichmann2022ApJ...939...43E}
{Eichmann}, B., {Oikonomou}, F., {Salvatore}, S., {Dettmar}, R.-J., \& {Tjus},
  J.~B. 2022, \apj, 939, 43

\bibitem[{{Erroz-Ferrer} {et~al.}(2015){Erroz-Ferrer}, {Knapen}, {Leaman},
  {Cisternas}, {Font}, {Beckman}, {Sheth}, {Mu{\~n}oz-Mateos},
  {D{\'\i}az-Garc{\'\i}a}, {Bosma}, {Athanassoula}, {Elmegreen}, {Ho}, {Kim},
  {Laurikainen}, {Martinez-Valpuesta}, {Meidt}, \&
  {Salo}}]{Erroz-Ferrer2015MNRAS.451.1004E}
{Erroz-Ferrer}, S., {Knapen}, J.~H., {Leaman}, R., {et~al.} 2015, \mnras, 451,
  1004

\bibitem[{{Esquej} {et~al.}(2014){Esquej}, {Alonso-Herrero},
  {Gonz{\'a}lez-Mart{\'\i}n}, {H{\"o}nig}, {Hern{\'a}n-Caballero}, {Roche},
  {Ramos Almeida}, {Mason}, {D{\'\i}az-Santos}, {Levenson}, {Aretxaga},
  {Rodr{\'\i}guez Espinosa}, \& {Packham}}]{Esquej2014ApJ...780...86E}
{Esquej}, P., {Alonso-Herrero}, A., {Gonz{\'a}lez-Mart{\'\i}n}, O., {et~al.}
  2014, \apj, 780, 86

\bibitem[{{Gardner} \& {Done}(2017)}]{Gardner2017MNRAS.470.3591G}
{Gardner}, E. \& {Done}, C. 2017, \mnras, 470, 3591

\bibitem[{{Glauch} {et~al.}(2023){Glauch}, {Kheirandish}, {Kontrimas}, {Liu},
  \& {Niederhausen}}]{IceCube2023arXiv230800024G}
{Glauch}, T., {Kheirandish}, A., {Kontrimas}, T., {Liu}, Q., \& {Niederhausen},
  H. 2023, arXiv e-prints, arXiv:2308.00024

\bibitem[{{Goswami}(2023)}]{IceCube2023arXiv230715349G}
{Goswami}, S. 2023, arXiv e-prints, arXiv:2307.15349

\bibitem[{{Guti{\'e}rrez} {et~al.}(2021){Guti{\'e}rrez}, {Vieyro}, \&
  {Romero}}]{Gutierrez2021A&A...649A..87G}
{Guti{\'e}rrez}, E.~M., {Vieyro}, F.~L., \& {Romero}, G.~E. 2021, \aap, 649,
  A87

\bibitem[{{H{\"o}nig} {et~al.}(2014){H{\"o}nig}, {Watson}, {Kishimoto}, \&
  {Hjorth}}]{Honing2014Natur.515..528H}
{H{\"o}nig}, S.~F., {Watson}, D., {Kishimoto}, M., \& {Hjorth}, J. 2014, \nat,
  515, 528

\bibitem[{{IceCube Collaboration} {et~al.}(2022){IceCube Collaboration},
  {Abbasi}, {Ackermann}, {Adams}, {Aguilar}, {Ahlers}, {Ahrens}, {Alameddine},
  {Alispach}, {Alves}, {Amin}, {Andeen}, {Anderson}, {Anton}, {Arg{\"u}elles},
  {Ashida}, {Axani}, {Bai}, {Balagopal}, {Barbano}, {Barwick}, {Bastian},
  {Basu}, {Baur}, {Bay}, {Beatty}, {Becker}, {Becker Tjus}, {Bellenghi},
  {Benzvi}, {Berley}, {Bernardini}, {Besson}, {Binder}, {Bindig}, {Blaufuss},
  {Blot}, {Boddenberg}, {Bontempo}, {Borowka}, {B{\"o}ser}, {Botner},
  {B{\"o}ttcher}, {Bourbeau}, {Bradascio}, {Braun}, {Brinson}, {Bron},
  {Brostean-Kaiser}, {Browne}, {Burgman}, {Burley}, {Busse}, {Campana},
  {Carnie-Bronca}, {Chen}, {Chen}, {Chirkin}, {Choi}, {Clark}, {Clark},
  {Classen}, {Coleman}, {Collin}, {Conrad}, {Coppin}, {Correa}, {Cowen},
  {Cross}, {Dappen}, {Dave}, {de Clercq}, {Delaunay}, {Delgado L{\'o}pez},
  {Dembinski}, {Deoskar}, {Desai}, {Desiati}, {de Vries}, {de Wasseige}, {de
  With}, {Deyoung}, {Diaz}, {D{\'\i}az-V{\'e}lez}, {Dittmer}, {Dujmovic},
  {Dunkman}, {Duvernois}, {Dvorak}, {Ehrhardt}, {Eller}, {Engel}, {Erpenbeck},
  {Evans}, {Evenson}, {Fan}, {Fazely}, {Fedynitch}, {Feigl}, {Fiedlschuster},
  {Fienberg}, {Filimonov}, {Finley}, {Fischer}, {Fox}, {Franckowiak},
  {Friedman}, {Fritz}, {F{\"u}rst}, {Gaisser}, {Gallagher}, {Ganster},
  {Garcia}, {Garrappa}, {Gerhardt}, {Ghadimi}, {Glaser}, {Glauch},
  {Gl{\"u}senkamp}, {Goldschmidt}, {Gonzalez}, {Goswami}, {Grant},
  {Gr{\'e}goire}, {Griswold}, {G{\"u}nther}, {Gutjahr}, {Haack}, {Hallgren},
  {Halliday}, {Halve}, {Halzen}, {Hanson}, {Hardin}, {Harnisch}, {Haungs},
  {Hebecker}, {Helbing}, {Henningsen}, {Hettinger}, {Hickford}, {Hignight},
  {Hill}, {Hill}, {Hoffman}, {Hoffmann}, {Hokanson-Fasig}, {Hoshina}, {Huang},
  {Huber}, {Huber}, {Hultqvist}, {H{\"u}nnefeld}, {Hussain}, {Hymon}, {in},
  {Iovine}, {Ishihara}, {Jansson}, {Japaridze}, {Jeong}, {Jin}, {Jones},
  {Kang}, {Kang}, {Kang}, {Kappes}, {Kappesser}, {Kardum}, {Karg}, {Karl},
  {Karle}, {Katz}, {Kauer}, {Kellermann}, {Kelley}, {Kheirandish}, {Kin},
  {Kintscher}, {Kiryluk}, {Klein}, {Koirala}, {Kolanoski}, {Kontrimas},
  {K{\"o}pke}, {Kopper}, {Kopper}, {Koskinen}, {Koundal}, {Kovacevich},
  {Kowalski}, {Kozynets}, {Kun}, {Kurahashi}, {Lad}, {Lagunas Gualda},
  {Lanfranchi}, {Larson}, {Lauber}, {Lazar}, {Lee}, {Leonard},
  {Leszczy{\'n}ska}, {Li}, {Lincetto}, {Liu}, {Liubarska}, {Lohfink}, {Lozano
  Mariscal}, {Lu}, {Lucarelli}, {Ludwig}, {Luszczak}, {Lyu}, {Ma}, {Madsen},
  {Mahn}, {Makino}, {Mancina}, {Mari{\c{s}}}, {Martinez-Soler}, {Maruyama},
  {Mase}, {McElroy}, {McNally}, {Mead}, {Meagher}, {Mechbal}, {Medina},
  {Meier}, {Meighen-Berger}, {Micallef}, {Mockler}, {Montaruli}, {Moore},
  {Morse}, {Moulai}, {Naab}, {Nagai}, {Nahnhauer}, {Naumann}, {Necker},
  {Nguyen}, {Niederhausen}, {Nisa}, {Nowicki}, {Nygren}, {Obertack},
  {Pollmann}, {Oehler}, {Oeyen}, {Olivas}, {O'Sullivan}, {Pandya}, {Pankova},
  {Park}, {Parker}, {Paudel}, {Paul}, {P{\'e}rez de Los Heros}, {Peters},
  {Peterson}, {Philippen}, {Pieper}, {Pittermann}, {Pizzuto}, {Plum},
  {Popovych}, {Porcelli}, {Prado Rodriguez}, {Price}, {Pries}, {Przybylski},
  {Rack-Helleis}, {Raissi}, {Rameez}, {Rawlins}, {Rea}, {Rehman},
  {Reichherzer}, {Reimann}, {Renzi}, {Resconi}, {Reusch}, {Rhode}, {Richman},
  {Riedel}, {Roberts}, {Robertson}, {Roellinghoff}, {Rongen}, {Rott}, {Ruhe},
  {Ryckbosch}, {Rysewyk Cantu}, {Safa}, {Saffer}, {Sanchez Herrera},
  {Sandrock}, {Sandroos}, {Santander}, {Sarkar}, {Sarkar}, {Satalecka},
  {Schaufel}, {Schieler}, {Schindler}, {Schmidt}, {Schneider}, {Schneider},
  {Schr{\"o}der}, {Schumacher}, {Schwefer}, {Sclafani}, {Seckel}, {Seunarine},
  {Sharma}, {Shefali}, {Silva}, {Skrzypek}, {Smithers}, {Snihur},
  {Soedingrekso}, {Soldin}, {Spannfellner}, {Spiczak}, {Spiering},
  {Stachurska}, {Stamatikos}, {Stanev}, {Stein}, {Stettner}, {Steuer},
  {Stezelberger}, {Stokstad}, {St{\"u}rwald}, {Stuttard}, {Sullivan},
  {Taboada}, {Ter-Antonyan}, {Tilav}, {Tischbein}, {Tollefson}, {T{\"o}nnis},
  {Toscano}, {Tosi}, {Trettin}, {Tselengidou}, {Tung}, {Turcati}, {Turcotte},
  {Turley}, {Twagirayezu}, {Ty}, {Unland Elorrieta}, {Valtonen-Mattila},
  {Vandenbroucke}, {van Eijndhoven}, {Vannerom}, {van Santen}, {Verpoest},
  {Walck}, {Watson}, {Weaver}, {Weigel}, {Weindl}, {Weiss}, {Weldert}, {Wendt},
  {Werthebach}, {Weyrauch}, {Whitehorn}, {Wiebusch}, {Williams}, {Wolf},
  {Woschnagg}, {Wrede}, {Wulff}, {Xu}, {Yanez}, {Yoshida}, {Yu}, {Yuan},
  {Zhangan}, \& {Zhelnin}}]{IceCube2022Sci...378..538I}
{IceCube Collaboration}, {Abbasi}, R., {Ackermann}, M., {et~al.} 2022, Science,
  378, 538

\bibitem[{{Inoue} \& {Doi}(2014)}]{Inoue2014PASJ...66L...8I}
{Inoue}, Y. \& {Doi}, A. 2014, \pasj, 66, L8

\bibitem[{{Inoue} \& {Doi}(2018)}]{Inoue2018ApJ...869..114I}
{Inoue}, Y. \& {Doi}, A. 2018, \apj, 869, 114

\bibitem[{{Inoue} {et~al.}(2020){Inoue}, {Khangulyan}, \&
  {Doi}}]{Inoue2020ApJ...891L..33I}
{Inoue}, Y., {Khangulyan}, D., \& {Doi}, A. 2020, \apjl, 891, L33

\bibitem[{{Inoue} {et~al.}(2019){Inoue}, {Khangulyan}, {Inoue}, \&
  {Doi}}]{Inoue2019ApJ...880...40I}
{Inoue}, Y., {Khangulyan}, D., {Inoue}, S., \& {Doi}, A. 2019, \apj, 880, 40

\bibitem[{{Khangulyan} {et~al.}(2014){Khangulyan}, {Aharonian}, \&
  {Kelner}}]{Khangulyan2014ApJ...783..100K}
{Khangulyan}, D., {Aharonian}, F.~A., \& {Kelner}, S.~R. 2014, \apj, 783, 100

\bibitem[{{Kishimoto} {et~al.}(2022){Kishimoto}, {Anderson}, {ten Brummelaar},
  {Farrington}, {Antonucci}, {H{\"o}nig}, {Millour}, {Tristram}, {Weigelt},
  {Sturmann}, {Sturmann}, {Schaefer}, \&
  {Scott}}]{Kishimoto2022ApJ...940...28K}
{Kishimoto}, M., {Anderson}, M., {ten Brummelaar}, T., {et~al.} 2022, \apj,
  940, 28

\bibitem[{{Kishimoto} {et~al.}(2013){Kishimoto}, {H{\"o}nig}, {Antonucci},
  {Millan-Gabet}, {Barvainis}, {Millour}, {Kotani}, {Tristram}, \&
  {Weigelt}}]{Kishimoto2013ApJ...775L..36K}
{Kishimoto}, M., {H{\"o}nig}, S.~F., {Antonucci}, R., {et~al.} 2013, \apjl,
  775, L36

\bibitem[{{Lamastra} {et~al.}(2016){Lamastra}, {Fiore}, {Guetta}, {Antonelli},
  {Colafrancesco}, {Menci}, {Puccetti}, {Stamerra}, \&
  {Zappacosta}}]{Lamastra2016A&A...596A..68L}
{Lamastra}, A., {Fiore}, F., {Guetta}, D., {et~al.} 2016, \aap, 596, A68

\bibitem[{{Li} {et~al.}(2022){Li}, {Feng}, {Liu}, {Bai}, {Li}, {Lu}, {Wang},
  {Huang}, \& {Zhang}}]{Li2022ApJ...936...75L}
{Li}, S.-S., {Feng}, H.-C., {Liu}, H.~T., {et~al.} 2022, \apj, 936, 75

\bibitem[{{Lin} {et~al.}(1993){Lin}, {Bertsch}, {Dingus}, {Fichtel}, {Hartman},
  {Hunter}, {Kanbach}, {Kniffen}, {Mattox}, {Mayer-Hasselwander}, {Michelson},
  {von Montigny}, {Nolan}, {Schneid}, {Sreekumar}, \&
  {Thompson}}]{Lin1993ApJ...416L..53L}
{Lin}, Y.~C., {Bertsch}, D.~L., {Dingus}, B.~L., {et~al.} 1993, \apjl, 416, L53

\bibitem[{{Liska} {et~al.}(2022){Liska}, {Musoke}, {Tchekhovskoy}, {Porth}, \&
  {Beloborodov}}]{Liska2022ApJ...935L...1L}
{Liska}, M.~T.~P., {Musoke}, G., {Tchekhovskoy}, A., {Porth}, O., \&
  {Beloborodov}, A.~M. 2022, \apjl, 935, L1

\bibitem[{{Mahmoud} \& {Done}(2020)}]{Mahmoud2020MNRAS.491.5126M}
{Mahmoud}, R.~D. \& {Done}, C. 2020, \mnras, 491, 5126

\bibitem[{{McEnery} {et~al.}(2019){McEnery}, {van der Horst}, {Dominguez},
  {Moiseev}, {Marcowith}, {Harding}, {Lien}, {Giuliani}, {Inglis}, {Ansoldi},
  {Stamerra}, {Manousakis}, {Strong}, {Bambi}, {Patricelli}, {Baring},
  {Barrio}, {Bastieri}, {Fields}, {Beacom}, {Beckmann}, {Bednarek}, {Rani},
  {Boggs}, {Bolotnikov}, {Cenko}, {Buckley}, {Grefenstette}, {Hui}, {Pittori},
  {Prescod-Weinstein}, {Shrader}, {Gouiffes}, {Kierans}, {Wilson-Hodge},
  {D'Ammando}, {Castro}, {Kocveski}, {Gasparrini}, {Thompson}, {Williams}, {De
  Angelis}, {Bernard}, {Digel}, {Morcuende}, {Charles}, {Bissaldi}, {Hays},
  {Ferrara}, {Bozzo}, {Grove}, {Wulf}, {Bottacini}, {Caroli}, {Kislat},
  {Oikonomou}, {Giordano}, {Longo}, {Fryer}, {Fukazawa}, {Georganopoulos}, {De
  Nolfo}, {Vianello}, {Kanbach}, {Younes}, {Blumer}, {Hartmann}, {Hernanz},
  {Takahashi}, {Li}, {Agudo}, {Moskalenko}, {Stumke}, {Grenier}, {Smith},
  {Rodi}, {Perkins}, {Gelfand}, {Holder}, {Knodlseder}, {Kopp}, {Lenain},
  {{\'A}lvarez}, {Metcalfe}, {Krizmanic}, {Stephen}, {Hewitt}, {Mitchell},
  {Harding}, {Tomsick}, {Racusin}, {Finke}, {Kargaltsev}, {Klimenko},
  {Krawczynski}, {Smith}, {Kubo}, {Di Venere}, {Marcotulli}, {Lommler},
  {Parker}, {Baldini}, {Foffano}, {Zampieri}, {Tibaldo}, {Petropoulou},
  {Ajello}, {Meyer}, {L{\'o}pez}, {McConnell}, {Boettcher}, {Cardillo},
  {Martinez}, {Kerr}, {Mazziotta}, {McEnery}, {Di Mauro}, {Wood}, {Meyer},
  {Briggs}, {De Becker}, {Lovellette}, {Doro}, {Sanchez-Conde}, {Moss},
  {Mizuno}, {Rib{\'o}}, {Nakazawa}, {Neilson}, {Auricchio}, {Omodei},
  {Oberlack}, {Ohno}, {Orlando}, {Otte}, {Coppi}, {Bloser}, {Zhang}, {Laurent},
  {Pohl}, {Prandini}, {Shawhan}, {Caputo}, {Campana}, {Rando}, {Woolf},
  {Johnson}, {Mignani}, {Walter}, {Ojha}, {da Silva}, {Dietrich}, {Funk},
  {Zane}, {Anton}, {Buson}, {Cutini}, {Saz Parkinson}, {Schirato}, {Griffin},
  {Kaufmann}, {Stawarz}, {Ciprini}, {Del Sordo}, {Jones}, {Guiriec}, {Tajima},
  {Cheung}, {The}, {Venters}, {Porter}, {Linden}, {Barres}, {Paliya},
  {Bozhilov}, {Vestrand}, {Tatischeff}, {Chen}, {Wang}, {Tanaka}, {Uhm},
  {Zhang}, {Zimmer}, {Zoglauer}, \& {Wadiasingh}}]{AMEGO2019BAAS...51g.245M}
{McEnery}, J., {van der Horst}, A., {Dominguez}, A., {et~al.} 2019, in Bulletin
  of the American Astronomical Society, Vol.~51, 245

\bibitem[{{Michiyama} {et~al.}(2023){Michiyama}, {Inoue}, \&
  {Doi}}]{Michiyama2023arXiv230615950M}
{Michiyama}, T., {Inoue}, Y., \& {Doi}, A. 2023, arXiv e-prints,
  arXiv:2306.15950

\bibitem[{{Michiyama} {et~al.}(2022){Michiyama}, {Inoue}, {Doi}, \&
  {Khangulyan}}]{2022ApJ...936L...1M}
{Michiyama}, T., {Inoue}, Y., {Doi}, A., \& {Khangulyan}, D. 2022, \apjl, 936,
  L1

\bibitem[{{Mundell} {et~al.}(1995){Mundell}, {Pedlar}, {Baum}, {O'Dea},
  {Gallimore}, \& {Brinks}}]{Mundell1995MNRAS.272..355M}
{Mundell}, C.~G., {Pedlar}, A., {Baum}, S.~A., {et~al.} 1995, \mnras, 272, 355

\bibitem[{{Murase} {et~al.}(2020){Murase}, {Kimura}, \&
  {M{\'e}sz{\'a}ros}}]{Murase2020PhRvL.125a1101M}
{Murase}, K., {Kimura}, S.~S., \& {M{\'e}sz{\'a}ros}, P. 2020, \prl, 125,
  011101

\bibitem[{{Oh} {et~al.}(2018){Oh}, {Koss}, {Markwardt}, {Schawinski},
  {Baumgartner}, {Barthelmy}, {Cenko}, {Gehrels}, {Mushotzky}, {Petulante},
  {Ricci}, {Lien}, \& {Trakhtenbrot}}]{Oh2018ApJS..235....4O}
{Oh}, K., {Koss}, M., {Markwardt}, C.~B., {et~al.} 2018, \apjs, 235, 4

\bibitem[{{Orlando} {et~al.}(2022){Orlando}, {Bottacini}, {Moiseev},
  {Bodaghee}, {Collmar}, {Ensslin}, {Moskalenko}, {Negro}, {Profumo}, {Digel},
  {Thompson}, {Baring}, {Bolotnikov}, {Cannady}, {Carini}, {Eberle}, {Grenier},
  {Harding}, {Hartmann}, {Herrmann}, {Kerr}, {Krivonos}, {Laurent}, {Longo},
  {Morselli}, {Philips}, {Sasaki}, {Shawhan}, {Shy}, {Skinner}, {Smith},
  {Stecker}, {Strong}, {Sturner}, {Tomsick}, {Wadiasingh}, {Woolf}, {Yates},
  {Ziock}, \& {Zoglauer}}]{GECCO2022JCAP...07..036O}
{Orlando}, E., {Bottacini}, E., {Moiseev}, A.~A., {et~al.} 2022, \jcap, 2022,
  036

\bibitem[{{Owen} {et~al.}(2021){Owen}, {Lee}, \&
  {Kong}}]{Owen2021MNRAS.506...52O}
{Owen}, E.~R., {Lee}, K.-G., \& {Kong}, A. K.~H. 2021, \mnras, 506, 52

\bibitem[{{Parker}(1955)}]{Parker1955ApJ...121..491P}
{Parker}, E.~N. 1955, \apj, 121, 491

\bibitem[{{Parker}(1966)}]{Parker1966ApJ...145..811P}
{Parker}, E.~N. 1966, \apj, 145, 811

\bibitem[{{Pedlar} {et~al.}(1993){Pedlar}, {Kukula}, {Longley}, {Muxlow},
  {Axon}, {Baum}, {O'Dea}, \& {Unger}}]{Pedlar1993MNRAS.263..471P}
{Pedlar}, A., {Kukula}, M.~J., {Longley}, D.~P.~T., {et~al.} 1993, \mnras, 263,
  471

\bibitem[{{Penston} \& {Perez}(1984)}]{Penston1984MNRAS.211P..33P}
{Penston}, M.~V. \& {Perez}, E. 1984, \mnras, 211, 33P

\bibitem[{{Peretti} {et~al.}(2023{\natexlab{a}}){Peretti}, {Lamastra},
  {Saturni}, {Ahlers}, {Blasi}, {Morlino}, \&
  {Cristofari}}]{Peretti2023arXiv230113689P}
{Peretti}, E., {Lamastra}, A., {Saturni}, F.~G., {et~al.} 2023{\natexlab{a}},
  arXiv e-prints, arXiv:2301.13689

\bibitem[{{Peretti} {et~al.}(2023{\natexlab{b}}){Peretti}, {Peron}, {Tombesi},
  {Lamastra}, {Ahlers}, \& {Saturni}}]{Peretti2023arXiv230303298P}
{Peretti}, E., {Peron}, G., {Tombesi}, F., {et~al.} 2023{\natexlab{b}}, arXiv
  e-prints, arXiv:2303.03298

\bibitem[{{Raginski} \& {Laor}(2016)}]{Raginski2016MNRAS.459.2082R}
{Raginski}, I. \& {Laor}, A. 2016, \mnras, 459, 2082

\bibitem[{{Ricci} \& {Trakhtenbrot}(2022)}]{Ricci2022arXiv221105132R}
{Ricci}, C. \& {Trakhtenbrot}, B. 2022, arXiv e-prints, arXiv:2211.05132

\bibitem[{{Robinson} {et~al.}(1994){Robinson}, {Vila-Vilaro}, {Axon}, {Perez},
  {Wagner}, {Baum}, {Boisson}, {Durret}, {Gonzalez-Delgado}, {Moles},
  {Masegosa}, {O'Brien}, {O'Dea}, {del Olmo}, {Pedlar}, {Penston}, {Perea},
  {Perez-Fournon}, {Rodriguez-Espinosa}, {Tadhunter}, {Terlevich}, {Unger}, \&
  {Ward}}]{Robinson1994A&A...291..351R}
{Robinson}, A., {Vila-Vilaro}, B., {Axon}, D.~J., {et~al.} 1994, \aap, 291, 351

\bibitem[{{Shapovalova} {et~al.}(2008){Shapovalova}, {Popovi{\'c}}, {Collin},
  {Burenkov}, {Chavushyan}, {Bochkarev}, {Ben{\'\i}tez}, {Dultzin},
  {Kova{\v{c}}evi{\'c}}, {Borisov}, {Carrasco}, {Le{\'o}n-Tavares}, {Mercado},
  {Valdes}, {Vlasuyk}, \& {Zhdanova}}]{Shapovalova2008A&A...486...99S}
{Shapovalova}, A.~I., {Popovi{\'c}}, L.~{\v{C}}., {Collin}, S., {et~al.} 2008,
  \aap, 486, 99

\bibitem[{{Takasao} {et~al.}(2018){Takasao}, {Tomida}, {Iwasaki}, \&
  {Suzuki}}]{Takasao2018ApJ...857....4T}
{Takasao}, S., {Tomida}, K., {Iwasaki}, K., \& {Suzuki}, T.~K. 2018, \apj, 857,
  4

\bibitem[{{Theios} {et~al.}(2016){Theios}, {Malkan}, \&
  {Ross}}]{Theios2016ApJ...822...45T}
{Theios}, R.~L., {Malkan}, M.~A., \& {Ross}, N.~R. 2016, \apj, 822, 45

\bibitem[{{Theureau} {et~al.}(2007){Theureau}, {Hanski}, {Coudreau}, {Hallet},
  \& {Martin}}]{Theureau2007A&A...465...71T}
{Theureau}, G., {Hanski}, M.~O., {Coudreau}, N., {Hallet}, N., \& {Martin},
  J.~M. 2007, \aap, 465, 71

\bibitem[{{Tombesi} {et~al.}(2010){Tombesi}, {Cappi}, {Reeves}, {Palumbo},
  {Yaqoob}, {Braito}, \& {Dadina}}]{Tombesi2010A&A...521A..57T}
{Tombesi}, F., {Cappi}, M., {Reeves}, J.~N., {et~al.} 2010, \aap, 521, A57

\bibitem[{{Tomsick} {et~al.}(2019){Tomsick}, {Zoglauer}, {Sleator}, {Lazar},
  {Beechert}, {Boggs}, {Roberts}, {Siegert}, {Lowell}, {Wulf}, {Grove},
  {Phlips}, {Brandt}, {Smale}, {Kierans}, {Burns}, {Hartmann}, {Leising},
  {Ajello}, {Fryer}, {Amman}, {Chang}, {Jean}, \& {von
  Ballmoos}}]{COSI2019BAAS...51g..98T}
{Tomsick}, J., {Zoglauer}, A., {Sleator}, C., {et~al.} 2019, in Bulletin of the
  American Astronomical Society, Vol.~51, 98

\bibitem[{{Ulvestad} {et~al.}(1998){Ulvestad}, {Roy}, {Colbert}, \&
  {Wilson}}]{Ulvestad1998ApJ...496..196U}
{Ulvestad}, J.~S., {Roy}, A.~L., {Colbert}, E. J.~M., \& {Wilson}, A.~S. 1998,
  \apj, 496, 196

\bibitem[{{Ulvestad} {et~al.}(2005){Ulvestad}, {Wong}, {Taylor}, {Gallimore},
  \& {Mundell}}]{Ulvestad2005AJ....130..936U}
{Ulvestad}, J.~S., {Wong}, D.~S., {Taylor}, G.~B., {Gallimore}, J.~F., \&
  {Mundell}, C.~G. 2005, \aj, 130, 936

\bibitem[{{Vila-Vilaro} {et~al.}(1995){Vila-Vilaro}, {Robinson}, {Perez},
  {Axon}, {Baum}, {Gonzalez-Delgado}, {Pedlar}, {Perez-Fournon}, {Perry}, \&
  {Tadhunter}}]{Vila-Vilaro1995A&A...302...58V}
{Vila-Vilaro}, B., {Robinson}, A., {Perez}, E., {et~al.} 1995, \aap, 302, 58

\bibitem[{{Williams} {et~al.}(2017){Williams}, {McHardy}, {Baldi}, {Beswick},
  {Argo}, {Dullo}, {Knapen}, {Brinks}, {Fenech}, {Mundell}, {Muxlow},
  {Panessa}, {Rampadarath}, \& {Westcott}}]{Williams2017MNRAS.472.3842W}
{Williams}, D.~R.~A., {McHardy}, I.~M., {Baldi}, R.~D., {et~al.} 2017, \mnras,
  472, 3842

\bibitem[{{Wilson} \& {Ulvestad}(1982)}]{Wilson1982ApJ...263..576W}
{Wilson}, A.~S. \& {Ulvestad}, J.~S. 1982, \apj, 263, 576

\bibitem[{{Yaqoob} {et~al.}(1989){Yaqoob}, {Warwick}, \&
  {Pounds}}]{Yaqoob1989MNRAS.236..153Y}
{Yaqoob}, T., {Warwick}, R.~S., \& {Pounds}, K.~A. 1989, \mnras, 236, 153

\bibitem[{{Yoshii} {et~al.}(2014){Yoshii}, {Kobayashi}, {Minezaki}, {Koshida},
  \& {Peterson}}]{Yoshii2014ApJ...784L..11Y}
{Yoshii}, Y., {Kobayashi}, Y., {Minezaki}, T., {Koshida}, S., \& {Peterson},
  B.~A. 2014, \apjl, 784, L11

\bibitem[{{Yuan} {et~al.}(2020){Yuan}, {Fausnaugh}, {Hoffmann}, {Macri},
  {Peterson}, {Riess}, {Bentz}, {Brown}, {Bont{\`a}}, {Davies}, {Rosa},
  {Ferrarese}, {Grier}, {Hicks}, {Onken}, {Pogge}, {Storchi-Bergmann}, \&
  {Vestergaard}}]{Yuan2020ApJ...902...26Y}
{Yuan}, W., {Fausnaugh}, M.~M., {Hoffmann}, S.~L., {et~al.} 2020, \apj, 902, 26

\bibitem[{{Zdziarski} {et~al.}(1996){Zdziarski}, {Johnson}, \&
  {Magdziarz}}]{Zdziarski1996MNRAS.283..193Z}
{Zdziarski}, A.~A., {Johnson}, W.~N., \& {Magdziarz}, P. 1996, \mnras, 283, 193

\end{thebibliography}
\end{document}